\documentstyle[12pt]{article}
\begin{document}
\thispagestyle{empty}
\newcommand{\p}[1]{(\ref{#1})}
\newcommand{\be}{\begin{equation}}
\newcommand{\ee}{\end{equation}}
\newcommand{\sect}[1]{\setcounter{equation}{0}\section{#1}}
\renewcommand{\theequation}{\thesection.\arabic{equation}}
\newcommand{\vs}[1]{\rule[- #1 mm]{0mm}{#1 mm}}
\newcommand{\hs}[1]{\hspace{#1mm}}
\newcommand{\Db}{{\overline D}}
\newcommand{\cD}{{\cal D}}
\newcommand{\cDb}{{\overline {\cal D}}}
\newcommand{\z}{z_{12}}
\newcommand{\tb}{{\bar\theta}_{12}}
\newcommand{\J}{{\cal J}}
\newcommand{\cH}{{\cal H}}
\newcommand{\cG}{{\cal G}}
\newcommand{\nn}{\nonumber}
\newcommand{\bea}{\begin{eqnarray}}
\newcommand{\eea}{\end{eqnarray}}
\newcommand{\wt}[1]{\widetilde{#1}}
\newcommand{\und}[1]{\underline{#1}}
\newcommand{\ov}[1]{\overline{#1}}
\newcommand{\sm}[2]{\frac{\mbox{\footnotesize #1}\vs{-2}}
		   {\vs{-2}\mbox{\footnotesize #2}}}
\newcommand{\prt}{\partial}
\newcommand{\eps}{\epsilon}

\newcommand{\R}{\mbox{\rule{0.2mm}{2.8mm}\hspace{-1.5mm} R}}
\newcommand{\Z}{Z\hspace{-2mm}Z}

\newcommand{\NP}[1]{Nucl.\ Phys.\ {\bf #1}}
\newcommand{\PL}[1]{Phys.\ Lett.\ {\bf #1}}
\newcommand{\NC}[1]{Nuovo Cimento {\bf #1}}
\newcommand{\CMP}[1]{Comm.\ Math.\ Phys.\ {\bf #1}}
\newcommand{\PR}[1]{Phys.\ Rev.\ {\bf #1}}
\newcommand{\PRL}[1]{Phys.\ Rev.\ Lett.\ {\bf #1}}
\newcommand{\MPL}[1]{Mod.\ Phys.\ Lett.\ {\bf #1}}
\newcommand{\BLMS}[1]{Bull.\ London Math.\ Soc.\ {\bf #1}}
\newcommand{\IJMP}[1]{Int.\ Jour.\ of\ Mod.\ Phys.\ {\bf #1}}
\newcommand{\JMP}[1]{Jour.\ of\ Math.\ Phys.\ {\bf #1}}
\newcommand{\LMP}[1]{Lett.\ in\ Math.\ Phys.\ {\bf #1}}

\renewcommand{\thefootnote}{\fnsymbol{footnote}}
\newpage
\setcounter{page}{0}
\pagestyle{empty}
\begin{flushright}
November 1997\\
SISSA 143/97/EP\\
solv-int/9711012
\end{flushright}
\vs{8}
\begin{center}
{\LARGE {\bf Coset approach to the  N=2 supersymmetric}}\\[0.6cm]
{\LARGE {\bf matrix GNLS hierarchies}}\\[1cm]

\vs{8}

{\large L. Bonora$^{a,1}$, S. Krivonos$^{b,2}$ and A. Sorin$^{b,3}$}
{}~\\
\quad \\
{\em ~$~^{(a)}$ International School for Advanced Studies (SISSA/ISAS),}\\
{\em Via Beirut 2, 34014 Trieste, Italy, and}\\
{\em INFN, Sezione di Trieste}\\
{\em {~$~^{(b)}$ Bogoliubov Laboratory of Theoretical Physics, JINR,}}\\
{\em 141980 Dubna, Moscow Region, Russia}~\quad\\

\end{center}
\vs{8}

\centerline{ {\bf Abstract}}
We discuss a large class of coset constructions of the N=2 $sl(n|n-1)$ affine 
superalgebra. We 
select admissible subalgebras, i.e. subalgebras that induce linear 
chiral/antichiral constraints on the coset supercurrents. We show that all 
the corresponding coset constructions lead to N=2 matrix GNLS hierarchies.
We develop an algorithm to compute the relative Hamiltonians and flows. 
We spell out completely the case of the N=2 $\widehat {sl}(3|2)$, 
which possesses four admissible subalgebras. The non--local second 
Hamiltonian structure of the N=2 matrix GNLS hierarchies is obtained via
Dirac procedure from the local N=2 $sl(n|n-1)$ affine superalgebra. 
We observe that to any second Hamiltonian structure with pure bosonic or 
pure fermionic superfield content there correspond two different N=2 matrix
GNLS hierarchies.
\vs{4}

\vfill
{\em E-Mail:\\
1) bonora@sissa.it\\
2) krivonos@thsun1.jinr.dubna.su\\
3) sorin@thsun1.jinr.dubna.su }
\newpage
\pagestyle{plain}
\renewcommand{\thefootnote}{\arabic{footnote}}
\setcounter{footnote}{0}

\section{Introduction}
In the last ten years there has been a great progress in the 
construction of new integrable hierarchies with extended $N\geq 2$
supersymmetry [1-16]. One of the main reasons of this success lies in the
fact that the second Hamiltonian structures of the KdV--type hierarchies
are extended superconformal algebras. This provides a way to classify 
supersymmetric KdV--type hierarchies by means of these superalgebras. 
The main steps in this direction
include the construction of the three different N=2 supersymmetric
KdV hierarchies [1-5] which possess the N=2 superconformal algebra as
their second Hamiltonian structures. Subsequently the $N=3$ \cite{n3kdv} , 
$N=4$ KdV [7-10] and N=2 Boussinesq hierarchies \cite{bouss}, related 
with $N=3,4$ super--Virasoro
and N=2 $W_3$ algebras, respectively, were found. The generalization
to the case of the hierarchies with N=2 $W_S$ $(S\geq 4)$ superalgebras as
the second Hamiltonian structures was initiated in \cite{n2w4} for the case of
N=2 $W_4$. Finally the two different Lax operators for two (out of three)
integrable hierarchies associated with N=2 $W_S$ have been proposed in 
\cite{DG} and \cite{BKS2}.

At the same time another series of integrable hierarchies with N=2 
supersymmetry,
the so called $(n,m)$-GNLS ones \cite{BKS1}, have been constructed. These 
hierarchies include only spin 1/2 chiral/antichiral superfields and 
possess non--local N=2
superalgebras as their second Hamiltonian structures \cite{BS}. 
One can try to construct new hierarchies by `joining' Lax operators of this
class with Lax operators of the previous (KdV) type.
The extensions of the KdV and Boussinesq hierarchies by junctions of their Lax
operators with GNLS one indeed have been found in \cite{IK,DG}. 
In \cite{ASS} it has been proved that the former extension is in fact 
gauge related to N=2 GNLS hierarchies.
Finally in \cite{BKS2} we have constructed the
matrix GNLS (MGNLS) N=2 supersymmetric hierarchies and proposed a large
class of different reductions of the N=2 matrix KP hierarchy.

As we said before, the N=2 $(n,m)$-GNLS hierarchies are generically
characterized by a non--local second Hamiltonian structure.
The appearance of systems with the non--local second Hamiltonian
structures seems ruin our hopes to classify the integrable hierarchies
via their second Hamiltonian structures, due to lack of the
classification of non--local algebras.
However, as was shown for the bosonic case in \cite{FK}, it is possible
to relate the non--local second Hamiltonian structure to a local one
via a coset construction. The
first examples in the case of N=2 supersymmetric
systems were elaborated in \cite{BK}. The aim of this letter is to show
that all $(k|l,m)$--MGNLS hierarchies (and $(l,m)$--GNLS as a particular
case for $k=1$) can be reproduced via the coset approach, starting from
the local N=2  ${\widehat {sl}}(n|n-1)$ superalgebra. The corresponding 
non--local superalgebras \cite{BS,BKS2} naturally appear in this scheme after 
applying Dirac reduction procedure. Thus, the classification problem 
reduces to the classification of {\it admissible} (see below) cosets of N=2 
${\widehat {sl}}(n|n-1)$ superalgebra.

The letter is organized as follows. In section 2 we briefly recall
the structure of N=2  ${\widehat {sl}}(n|n-1)$ superalgebra \cite{AIS}. 
The N=2 supercurrents $\J$ which span the ${\widehat {sl}}(n|n-1)$ 
superalgebra, satisfy nonlinear chirality constraints. 
To construct the cosets of N=2 ${\widehat {sl}}(n|n-1)$ with 
the chiral/antichiral superfields, we
find  the subalgebras $\cH_s$ for which the nonlinear chirality
constraints on the coset currents $\J_{{\widehat {sl}}(n|n-1)/\cH_s}$ 
reduce to the  linear
chiral/antichi\-r\-al ones on the shell of constraints $\J_{\cH_s}=0$.
We call these {\it admissible} cosets. 
In Section 3 we
demonstrate how the coset approach works in the simplest non-trivial
case of N=2 ${\widehat {sl}}(3|2)$ superalgebra. We derive
the non--local superalgebras for the coset supercurrents using the Dirac
procedure. Moreover, we propose
the general form of the admissible cosets of the 
N=2 ${\widehat {sl}}(n|n-1)$ superalgebra for generic $n$.  

\section{The N=2 affine superalgebra $sl(n|n-1)$  and its admissible 
subalgebras}
Here, we give a short account of the N=2  superalgebra 
${\widehat {sl}}(n|n-1)$ \cite{AIS}, which will be needed in the construction of
the N=2 super $(k|l,m)$--MGNLS hierarchies and 
their second Hamiltonian structures.

The N=2  superalgebra ${\widehat {sl}}(n|n-1)$ contains an equal number, 
precisely $2n(n-1)$, of fermionic and bosonic N=2 supercurrents obeying 
the proper 
covariant nonlinear chirality conditions. In reference to the superalgebra
${\widehat {sl}}(n|n-1)$, we use the following convention for the
fundamental representation: we divide any supermatrix $\J_{\alpha\beta}$
into four blocks; the top--left $n\times n$ and bottom-right 
$(n-1)\times (n-1)$ blocks
are made of fermionic supercurrents, while the bosonic
supercurrents are the entries of the top-right $n\times (n-1)$ and 
bottom-left $(n-1)\times n$ blocks.

We introduce the grading $d_{\alpha\beta} $ for the supercurrents 
$\J_{\alpha\beta}$ which is equal to 1 for the 
fermionic and 0 for the bosonic currents. 
Following \cite{AIS} we will also use the complex basis for N=2
${\widehat {sl}}(n|n-1)$ superalgebra, labeling its generators by $a$ and 
$\bar a$;
$a,{\bar a} = 1,2,\ldots ,\frac{1}{2}((2n-1)^2-1)$. The explicit
relation between these two bases is given by
\bea
\J_{\alpha\beta}=
\left( \begin{array}{cccccccccc}
 h_{\bar{1}}& & & & & & & &    \\
             & h_{\overline{2}}+h_{1}&  & & & &
\mbox{unbarred} & &    \\
             & &. &  & & & \mbox{indices} & &    \\
             & & &  & & & & &   \\
             & & & & h_{\overline{n-1}}+h_{n-2} & & & &    \\
             & & & & &  h_{{n-1}} & & &    \\
             & & & & & & h_{\overline{1}}+h_{1}  & &    \\
             & \mbox{barred} & & & & & &. & &   \\
             &  \mbox{indices} & & & & & & &   \\
             & & & & & & & &   h_{\overline{n-1}}+h_{n-1}
\end{array} \right) \; .
\label{jmn}
\eea
Barred (unbarred) indices stand for $\J_{\bar a}$ ($\J_a$), these generators
being written in a prescribed order.
The Poisson brackets\footnote{In what follows all the operators appearing in 
the r.h.s. 
of the Poisson brackets are evaluated at the point $Z_1$ and the derivatives 
are assumed to act freely to the right.} between the supercurrents 
$\J_a,\J_{\bar a}$ form the N=2  superalgebra ${\widehat {sl}}(n|n-1)$ 
at level $K$
\bea
\left\{ \J_{a}(Z_1) , \J_b(Z_2) \right\} & = & 
-\left( f_{ab}{}^c D\J_c +
 \frac{1}{K}(-1)^{(d_a+d_b)d_{\bar c}}
 f_{a{\bar c}}{}^d f_b{}^{{\bar c}e}\J_d\J_e \right)\delta (1,2)\; , \nn \\
\left\{ \J_{\bar a}(Z_1), \J_{\bar b}(Z_2)\right\} & = & \left(
 -f_{{\bar a}{\bar b}}{}^{\bar c}\Db \J_{\bar c}+
 \frac{1}{K}(-1)^{(d_{\bar a}+d_{\bar b})d_c}
 f_{{\bar a}c}{}^{\bar d} 
 f_{\bar b}{}^{c{\bar e}}\J_{\bar d}\J_{\bar e}\right) \delta (1,2) \; ,
\nn \\
\left\{ \J_a(Z_1), \J_{\bar b}(Z_2)\right\} & = & -\left( Kg_{a{\bar b}}D\Db
 +(-1)^{d_c}f_{a{\bar b}}{}^c\J_c\Db+
 f_{a{\bar b}}{}^{\bar c}D\J_{\bar c} \right .\nn \\
 & & \left. + \frac{1}{K} (-1)^{(d_a+d_{\bar b})d_{\bar c}}
f_{a{\bar c}}{}^{d}f_{\bar b}{}^{{\bar c}{\bar e}}
 \J_d\J_{\bar e} \right) \delta (1,2) \; , \label{alg}
\eea 
where $Z=(z,\theta,\bar\theta)$ is a coordinate of the N=2 superspace,
$\delta (1,2) =(\theta_1-\theta_2)({\bar \theta}_1-{\bar\theta}_2)
 \delta (z_1-z_2)$ is the N=2 superspace delta--function
and the nonlinear chirality constraints on the supercurrents have the
following form:
\be\label{constr1}
D\J_a-\frac{1}{2K}(-1)^{d_a}f_a{}^{bc}\J_b\J_c=0, \quad\;
\Db\J_{\bar a}+\frac{1}{2K}(-1)^{d_{\bar a}}
 f_{\bar a}{}^{{\bar b}{\bar c}}\J_{\bar b}\J_{\bar c}=0
\ee
(the summation over repeated indices is assumed). Here, 
$D$ and ${\overline D}$ are the N=2
supersymmetric fermionic covariant derivatives
\begin{equation}\label{DD}
D=\frac{\partial}{\partial\theta}
 -\frac{1}{2}\bar\theta\frac{\partial}{\partial z} , \quad
{\overline D}=\frac{\partial}{\partial\bar\theta}
 -\frac{1}{2}\theta\frac{\partial}{\partial z} ,\quad
\left\{ D,{\overline D} \right\}= -\frac{\partial}{\partial z}, \quad
\left\{ D,D \right\} = \left\{ {\overline D},{\overline D} \right\}= 0.
\end{equation}
The structure constants 
$f_{ab}{}^c, f_{{\bar a}{\bar b}}{}^{\bar c},f_{ab}{}^{\bar c}$ and Killing
metric $g_{a{\bar b}}$ are defined in \cite{AIS}.

Before going further, let us remark that the choice of the Cartan
currents $h_a,h_{\bar a}$ \p{jmn} is imposed by the requirement
that the nonlinear
covariant chirality conditions \p{constr1} become for them standard
linear chiral/antichiral ones: $Dh_a = \Db h_{\bar a} =0 \;.$

The main aim of these letter is to demonstrate that the N=2 $(k|l,m)$--MGNLS
hierarchies and their second Hamiltonian structures \cite{BKS2} can be
obtained via coset approach from the N=2  ${\widehat {sl}}(n|n-1)$ 
superalgebra.

The basic step in the construction of any coset $\cG/\cH$ is the choice of
the stability subalgebra $\cH$. Once this is done, one has to find the 
Hamiltonians $H_p$ which belong to the coset space, i.e.
commute with the currents $\J_{\cH}$ of the subalgebra $\cH$ and 
therefore give the trivial equations of motion for them. 
Just these Hamiltonians  will produce the equations
of motion for the coset currents $\J_{\cG/\cH}$ on the
constraint shell $\J_{\cH}=0$: 
\be\label{shell}
\frac{\partial}{\partial t_p} \J_{\cG/\cH} = \left.\left\{
     H_p, \J_{\cG/\cH}\right\}\right|_{\J_{\cH}=0}\; .
\ee
If the constraints $\J_{\cH}=0$ are second class, then one has to find
the Dirac brackets for the coset currents. They will provide
the second Hamiltonian structure for the system \p{shell}.

Let us recall that the first and second flow equations of the N=2
supersymmetric $(k|l,m)$--MGNLS hierarchy have the following form \cite{BKS2}
\bea
\frac{\partial}{\partial t_1}F & = & F' ,\quad 
  \frac{\partial}{\partial t_1}{\overline F}={\overline F}' ; \nn \\
\frac{\partial}{\partial t_2}F & = & F''+\frac{1}{2K^2}
      D(F{\overline F}\;\Db F),\quad
      \frac{\partial}{\partial t_2}{\overline F} = -{\overline F}''+
      \frac{1}{2K^2} \Db (( D{\overline F})F{\overline F}), 
      \label{mgnls}
\eea
where $F\equiv F_{Aa}(Z)$ and ${\overline F}\equiv {\overline F}_{bB}(Z)
\quad (A,B =1, \ldots ,k; a,b=1,\ldots , l+m)$ are chiral and antichiral
rectangular matrix-valued N=2 superfields,
\be
DF=0, \quad \Db\;{\overline F}=0, \label{fchir}
\ee
respectively. In \p{mgnls} the matrix product is understood. 
The matrix entries are bosonic 
superfields for $a=1,\ldots ,l$ and fermionic superfields for
$a=l+1,\ldots, l+m$. The parameter $K$ in the flow equations \p{mgnls}
could be set equal to 1 by suitably rescaling the superfield matrices $F$
and $ \overline F$.
But, as we will see in the next section, in the coset approach it coincides 
with the level of the  superalgebra we start from, therefore we prefer
to write it explicitly.

Thus, all the supercurrents in the N=2 $(k|l,m)$--MGNLS
hierarchies satisfy linear chiral/antichi\-r\-al conditions \p{fchir},
while in the N=2  ${\widehat {sl}}(n|n-1)$ superalgebra the supercurrents are
nonlinearly constrained \p{constr1}. Therefore our first aim is to find the 
{\it admissible} subalgebras $\cH$ of the N=2 ${\widehat {sl}}(n|n-1)$ 
superalgebra, i.e. those subalgebras 
for which the nonlinear chirality
constraints \p{constr1} for the coset supercurrents $\J_{{\cG}/{\cH}}$
reduce to the linear chiral/antichiral ones on the shell $\J_{\cH} = 0$. 
It is somewhat surprising that we can list (a large class of)
such subalgebras of N=2 ${\widehat {sl}}(n|n-1)$ with a very simple recipe.
In fact there are $2n-2$ admissible
subalgebras. They are identified by the supermatrices formed by extracting the
top--left $s\times s$
and bottom--right $(2n-1-s)\times(2n-1-s)$ blocks out of the supermatrix
$\J_{\alpha\beta}$ \p{jmn}:
\bea
\J_{\alpha\beta}=
\left( \begin{array}{c|c}
 &  \\
\J_{{\cH}_s}& \J_{\cG/{\cH}_s}     \\
 &  \\
{\scriptstyle s\times s} & {\scriptstyle  s\times (2n-s-1)} \\ \hline 
 &  \\
\J_{\cG/{\cH}_s} & \J_{{\cH}_s}   \\
 &  \\
{\scriptstyle (2n-s-1)\times s }& {\scriptstyle (2n-s-1)\times (2n-s-1)}
\end{array} \right)
\label{subgr}
\eea 
The currents $\J_{{\cH}_s}$ span the following N=2 superalgebras as
$s$ runs from 1 to $2n-2$:
\bea \label{sub1}
{\widehat {gl}}(s)\oplus {\widehat {sl}}(n-s|n-1)  ,& s=2r \leq n, \nn \\
{\widehat {gl}}(n|s-n)\oplus {\widehat {sl}}(2n-s-1)  ,& s=2r > n \quad ,
\eea
and
\bea \label{sub2}
{\widehat {sl}}(s)\oplus {\widehat {gl}}(n-s|n-1)  ,& s=2r+1 \leq n,\nn \\
{\widehat {sl}}(n|s-n)\oplus {\widehat {gl}}(2n-s-1)  ,& s=2r+1 > n \quad .
\eea

Now we can make three statements.
First, the supercurrents in the top--left $s\times s $ and bottom--right 
$(2n-s-1)\times (2n-s-1)$  
blocks \p{subgr} do not form N=2 subalgebras separately.
In fact in order to expose the structure \p{sub1},\p{sub2}, a suitable 
recombination of the Cartan currents is necessary.
Second, the general formulae for Poisson brackets \p{alg}
and constraints \p{constr1} on the currents of the N=2 
${\widehat {sl}}(n|n-1)$ superalgebra provide the corresponding expressions 
for the
subalgebras \p{sub1} and \p{sub2} without any modification. 
All we have to do
to get the Poisson brackets and the constraints on 
the currents for the N=2 superalgebras from the list \p{sub1},\p{sub2} at
a given value of $s$, is to restrict the indices of the supercurrent
$\J_{\alpha\beta}$ \p{jmn} to run whithin the corresponding top--left and
bottom--right blocks. For example, the currents $\J_{11},\J_{\alpha\beta} \quad
(\alpha,\beta\geq 2)$ form the superalgebra ${\widehat {gl}}(n-1|n-1)$, 
while the
currents 
$\J_{(2n-1)(2n-1)},\J_{\alpha\beta} \quad (\alpha,\beta\leq (2n-2))$ span
${\widehat {gl}}(n|n-2)$
(with constraints \p{constr1} taken into account). The last statement we have to
make 
is that the constraints \p{constr1} on the coset currents $\J_{\cG/{\cH}_s}$
are reduced to the pure chiral/antichiral ones if we put $\J_{{\cH}_s}=0$
for all subalgebras from \p{sub1},\p{sub2}.
All these statements can be easily proved after rewriting the
Poisson brackets \p{alg} and constraints \p{constr1} 
for the supercurrents \p{jmn} in the fundamental representation. We will
not present the proof here.

After identifying the admissible subalgebras, in order 
to construct the cosets of N=2 ${\widehat {sl}}(n|n-1)$ with respect to 
any given subalgebra  $\cH_s$ \p{sub1} or \p{sub2},
we have to find the Hamiltonian densities which commute with all
supercurrents from $\cH_s$. They will include in general arbitrary parameters.
Next, using \p{shell}, we find the equations of motion for the
coset supercurrents $\J_{\cG/\cH_s}$ (with the condition $\J_{\cH_s}=0$) and 
then fix the parameters in the Hamiltonians in order to recover integrability. 
The first two steps are purely
technical ones, but in practice the straightforward calculations 
become quickly very cumbersome because a lot supercurrents are involved. 
A useful way to simplify the calculations
is to define new fermionic derivatives for all the coset supercurrents,
$\cD \J_{\cG/\cH_s}$ and $\cDb \J_{\cG/\cH_s}$, which are covariant with
respect to
$\cH_s$ (i.e. they transform to themselves with respect to the adjoint
action of the subalgebra $\cH_s$ at $\J_{\cH_s}= 0$). By means of covariant
derivatives it is therefore easy
to construct $\cH_s$--invariants. In fact, as we will see, 
all currents from $\cH_s$  
appear in the Hamiltonians only through these derivatives. 

On the other hand proving integrability is a more difficult task.
As usual, either we are able to construct the relevant Lax operators or  
we are able to show that the constructed systems coincide with known 
integrable ones.
Our aim here is not to prove integrability in general, but to demonstrate,
for all admissible subalgebras
of the N=2 ${\widehat {sl}}(3|2)$ superalgebra,
that we can construct the corresponding cosets and that they do reproduce 
the integrable
N=2 supersymmetric $(k|l,m)$--MGNLS hierarchies \cite{BKS2}.
We also suggest that the same construction holds for the generic case of 
N=2 ${\widehat {sl}}(n|n-1)$ superalgebra.

\sect{N=2 ${\widehat {sl}}(3|2)$ superalgebra and MGNLS hierarchies}
In this section we consider the cosets of the N=2 ${\widehat {sl}}(3|2)$
superalgebra with respect to its four admissible subalgebras from the list 
\p{sub1},\p{sub2} and 
show that these cosets give rise to the N=2 MGNLS hierarchies.

\subsection{N=2 ${\widehat {sl}}(3|2)/{\widehat {gl}}(2|2)$ coset 
construction}
The first subalgebra $\cH_1$ from the list \p{sub1},\p{sub2} for the 
N=2 ${\widehat {sl}}(3|2)$ 
superalgebra is the N=2  superalgebra ${\widehat {gl}}(2|2)$. Its currents 
and the N=2 ${\widehat {sl}}(3|2)/{\widehat {gl}}(2|2)$ coset currents are 
extracted from $\J_{\alpha\beta}$ \p{jmn},
\be
\J_{\cH_1}=
\left( \begin{array}{c|cccc}
 h_{\bar 1} & & & &  \\ \hline
 & h_{\bar 2}+h_1 & j_{23}&j_{24} &j_{25}     \\
 & j_{32}&  h_2 & j_{34} &j_{35} \\
 & j_{42} & j_{43} & h_{\bar 1}+h_1 & j_{45} \\
 & j_{52} & j_{53} & j_{54}& h_{\bar 2}+h_2 
\end{array} \right) \; , \;
\J_{{\widehat {sl}}(3|2)/{\cH_1}}=
\left( \begin{array}{c|cccc}
  &j_{12} &j_{13} &j_{14} & j_{15} \\ \hline
j_{21} &  & & &     \\
j_{31} &  & & &     \\
j_{41} &  & & &     \\
j_{51} &  & & &     \\
\end{array} \right),
\label{coset1}
\ee
respectively.
The coset currents $j_{12},j_{13},j_{21},j_{31}$ are fermionic while the
remaining ones are bosonic. It is very convenient to group these 
supercurrents into a row $F$ and a column ${\overline F}$:
\be
F=\left\{ j_{14},j_{15}, j_{12},j_{13} \right\} \quad ,
{\overline F}=\left\{
\begin{array}{c}
j_{41}\\j_{41}\\-j_{21}\\-j_{31}
\end{array} \right\} . \label{mat1}
\ee
To construct the first two Hamiltonians $H_{1,2}$ which commute 
with 
all the currents of N=2 ${\widehat {gl}}(2|2)$, and therefore belong to 
the coset space ${\widehat {sl}}(3|2)/{\widehat {gl}}(2|2)$ (and provide 
the trivial equations for all the
currents of ${\widehat {gl}}(2|2): \partial \J_{{\widehat {gl}}(2|2)}/\partial 
t_{1,2} =0
$), we need to introduce the covariant derivatives for the coset currents.
They can be defined as follows:
\bea
\cDb j_{12} & = & \Db j_{12}-\frac{1}{K}\left( j_{12}h_{\bar 1}-
 j_{13}j_{32}+j_{14}j_{42}+j_{15}j_{52}\right) \; , \;
\cDb j_{14}  =  \Db j_{14}+\frac{1}{K}\left( j_{14}h_{\bar 1}+
  j_{15}j_{54}\right) \; , \nn \\
\cDb j_{13} & = & \Db j_{13}-\frac{1}{K}\left( j_{13}h_{\bar 1}+
 j_{13}h_{\bar 2}+j_{14}j_{43}+j_{15}j_{53} \right)\; , \;
\cDb j_{15}  =  \Db j_{15}+\frac{1}{K}\left( j_{15}h_{\bar 1}+
 j_{15}h_{\bar 2} \right)\; , \nn \\
\cD j_{21} & = & D j_{21}-\frac{1}{K}\left( h_1 j_{21}+
 j_{23}j_{31}-j_{24}j_{41}-j_{25}j_{51}\right) \; , \;
\cD j_{41}  =  D j_{41}+\frac{1}{K} j_{45} j_{51} \; , \nn \\
\cD j_{31} & = & D j_{31}-\frac{1}{K}\left( h_1 j_{31}+
 h_2j_{31}-j_{34}j_{41}-j_{35}j_{51}\right) \; , \;
\cD j_{51}  =  D j_{51}-\frac{1}{K} h_1 j_{51} \; . \label{cd1}
\eea
The expressions for the covariant derivatives \p{cd1} look complicated, but
their construction is straightforward.
The first term is always the fermionic derivative of the given current; then 
we have some additional terms, which can be easily constructed using the basic
requirement that the Poisson brackets of covariant derivatives with
the currents of ${\widehat {gl}}(2|2)$ contain only these derivatives and
${\widehat {gl}}(2|2)$ currents.

Now we are ready to construct the first two Hamiltonians:
\be
H_1 = -\frac{1}{K} \int\! dZ \mbox{ tr} \left(F{\overline F}\right) \; , \;
H_2 =  -\frac{1}{K} \int\! dZ \mbox { tr} 
\left(( \cD {\overline F})\cDb F  -
  \frac{1}{2K^2} F{\overline F} F {\overline F} \right) \; ,
\label{h12}
\ee
(here the matrix multiplication is assumed).
To get the equations of motion for the coset superfields we use eq.\p{shell}.  
One can check that the resulting first and second flow equations 
coincide with the corresponding equations of the $(1|2,2)$--MGNLS
hierarchy \p{mgnls}. For the reader's convenience we present eq.\p{constr1}
in explicit form fo the coset supercurrents \p{mat1}, in this particular case,
\bea
Dj_{12}& = & -\frac{1}{K}h_1j_{12} , \;
Dj_{13} =  -\frac{1}{K}\left((h_1+h_2)j_{13}-j_{12}j_{23}\right),\;
Dj_{14}  =  \frac{1}{K}\left(j_{12}j_{24}+j_{13}j_{34}\right) , \nn \\
Dj_{15}  & =& 
-\frac{1}{K}\left( h_1j_{15}-j_{12}j_{25}-j_{13}j_{35}-
   j_{14}j_{45}\right) , \;
\Db j_{21}  =  \frac{1}{K}h_{\bar 1}j_{21} , \;
\Db j_{31} =\frac{1}{K}\left( (h_{\bar 1}+h_{\bar 2})j_{31}+
        j_{21}j_{32}\right) , \nn \\
\Db j_{41}  & = & \frac{1}{K}\left(h_{\bar 1}j_{41}+
   j_{21}j_{42}+j_{31}j_{43}\right) , \;
\Db j_{51}=\frac{1}{K}\left( (h_{\bar 1}+h_{\bar 2})j_{51}+
  j_{21}j_{52}+j_{31}j_{53}+j_{41}j_{54}\right). \label{ccc1}
\eea
After putting all supercurrents $\J_{\cH_1}$, \p{coset1}, equal to
zero, they become the linear chiral/antichiral ones, 
$ D F = \Db {\overline F} =0$.
Thus we end up with the chiral/antichiral matrix superfields 
$F, {\overline F}$ \p{mat1} which obey the equations \p{mgnls}. Therefore
we have succeeded in reproducing the first two flows of the
$(1|2,2)$--MGNLS hierarchy in the framework of the N=2
coset space ${\widehat {sl}}(3|2)/{\widehat {gl}}(2|2)$.

Let us close this subsection with some comments which will also apply to
the next subsections.

First of all we would like to repeat that here and in the coming subsections,
we quotient only with admissible subalgebras $\cH_s$ \p{sub1},\p{sub2} of 
the N=2 ${\widehat {sl}}(3|2)$ superalgebra. This is what guarantees
that in the corresponding coset spaces the non--linear chirality constraints 
\p{constr1} are reduced to the linear chiral/antichiral ones.

Second, we remark that using the matrix $F$ and
$\overline F$ \p{mat1} together with the covariant derivatives \p{cd1}
greatly reduces the number of possible terms in the Hamiltonians
 \p{h12}. Without introducing these covariant derivatives even 
$H_2$ would contain a huge number of terms -- remember that the initial system 
has 24 supercurrents!

Third, it immediately follows from our construction that the
non--local second Hamiltonian structure for the N=2 $(1|2,2)$--MGNLS 
($(2,2)$--GNLS) hierarchy \cite{BS}
can be obtained from the {\it local} N=2  ${\widehat {sl}}(3|2)$ superalgebra 
through the Dirac procedure. This can be done following 
the lines of \cite{BK}.

Finally, let us remark that the two terms in the Hamiltonian
$H_2$ \p{h12} commute separately with the all currents of the N=2 
superalgebra ${\widehat {gl}}(2|2)$, therefore they would allow for two
arbitrary coefficients in the equations of motion. We have fixed
these coefficients in order to reproduce the known $(1|2,2)$--MGNLS hierarchy. 

\subsection{N=2 
${\widehat {sl}}(3|2)/{\widehat {gl}}(2)\oplus {\widehat {sl}}(2|1)$ coset
construction}
The currents spanned the next N=2 subalgebra 
$\cH_2={\widehat {gl}}(2)\oplus {\widehat {sl}}(2|1)$ from the list
\p{sub1},\p{sub2} 
and the currents of the coset space
${\widehat {sl}}(3|2)/{\widehat {gl}}(2)\oplus{\hat  sl}(2|1)$ fill the
following blocks in $\J_{\alpha\beta}$ \p{jmn}:
\be
\J_{\cH_2}=
\left( \begin{array}{cc|ccc}
 h_{\bar 1} &j_{12} & & &  \\ 
 j_{21} & h_{\bar 2}+h_1 & & &     \\ \hline
 & &  h_2 & j_{34} &j_{35} \\
 & & j_{43} & h_{\bar 1}+h_1 & j_{45} \\
 &  & j_{53} & j_{54}& h_{\bar 2}+h_2 
\end{array} \right) \; , \;
\J_{{\widehat {sl}}(3|2)/{\cH_2}}=
\left( \begin{array}{cc|ccc}
  & &j_{13} &j_{14} & j_{15} \\ 
  & &j_{23} &j_{24} & j_{25}    \\ \hline
j_{31} & j_{32} & & &     \\
j_{41} & j_{42} & & &     \\
j_{51} & j_{52} & & &     \\
\end{array} \right),
\label{coset2}
\ee
respectively.
Now, among the coset currents, the fermionic ones are 
$j_{13},j_{23},j_{31},j_{32}$, while the
eight remaining ones are bosonic. Let us again arrange these supercurrents into
two matrices -- $F$ and ${\overline F}$:
\be
F=\left\{ \begin{array}{ccc}
  j_{14} & j_{15} &j_{13} \\
 j_{24} & j_{25} & j_{23} 
\end{array}
 \right\}, \quad 
{\overline F}=\left\{
\begin{array}{cc}
j_{41}& j_{42}\\
j_{51}& j_{52}\\
-j_{31}& -j_{32}
\end{array} \right\} . \label{mat2}
\ee
After introducing the covariant derivatives for the coset currents,
which we define as:
\bea
\cDb j_{13} & = & \Db j_{13}-\frac{1}{K}\left( j_{13}h_{\bar 2}+
 j_{13}h_{\bar 1}+j_{14}j_{43}+j_{15}j_{53}\right) \; , \;
\cDb j_{14}  =  \Db j_{14}+\frac{1}{K}\left( j_{14}h_{\bar 1}+
  j_{15}j_{54} \right)\; , \nn \\
\cDb j_{15} & = & \Db j_{15}+\frac{1}{K}\left( j_{15}h_{\bar 1}+
  j_{15}h_{\bar 2}\right) \; , \;
\cDb j_{23}  =  \Db j_{23}-\frac{1}{K}\left( j_{23}h_{\bar 2}+
 j_{13}j_{21}+j_{24}j_{43}+j_{25}j_{53} \right)\; , \nn \\
\cDb j_{24} & = & \Db j_{24}+\frac{1}{K}\left( j_{14}j_{21}+
  j_{25}j_{53} \right)\; , \;
\cDb j_{25}  =  \Db j_{25}+\frac{1}{K}\left( j_{15}j_{21}+
  j_{25}h_{\bar 2}\right) \; , \nn \\
\cD j_{31} & = & D j_{31}-\frac{1}{K}\left( h_1 j_{31}+
 h_2j_{31}-j_{34}j_{41}-j_{35}j_{51}\right) \; , \;
\cD j_{41}  =  D j_{41}+\frac{1}{K} j_{45} j_{51}\;, \nn \\
\cD j_{32} & = & D j_{32}-\frac{1}{K}\left( h_2 j_{32}-
 j_{12}j_{31}-j_{34}j_{42}-j_{35}j_{52}\right) \; , \;
\cD j_{51}  =  D j_{51}-\frac{1}{K} h_1 j_{51}\;, \nn \\
\cD j_{42} & = & D j_{42}+\frac{1}{K} \left( h_1 j_{42}+
  j_{12}j_{41}+j_{45}j_{52} \right) \; , \;
\cD j_{52}  =  D j_{52}+\frac{1}{K} j_{12} j_{51} \; , \label{cd2}
\eea
one can check that the first two Hamiltonians \p{h12}
give rise to the first two flows, \p{mgnls}, of the N=2 
$(2|2,1)$--MGNLS hierarchy.
Once again, the matrix supercurrents $F, {\overline F}$  become 
chiral/antichiral ones
on the shell $\J_{{\widehat {gl}}(2)\oplus {\widehat {sl}}(2|1)}=0$.

Thus our second N=2 coset construction, 
${\widehat {sl}}(3|2)/{\widehat {gl}}(2)\oplus {\hat sl}(2|1)$, 
give rise to the N=2 $(2|2,1)$--MGNLS hierarchy. 

The N=2 $(2|2,1)$--MGNLS hierarchy we constructed here in the framework
of the coset approach is the unique example of matrix hierarchy with both
fermionic and bosonic superfields among the four coset constructions 
of the N=2 ${\widehat {sl}}(3|2)$ superalgebra. Therefore it is
instructive to explicitly construct the Poisson brackets for the coset 
matrix superfields $F$ and $\overline F$, \p{mat2}.
However, this cannot be done straightforwardly because 
the constraints $\J_{{\widehat {gl}}(2)\oplus {\widehat {sl}}(2|1)} =0$
are mixed first and second class. This can be immediately seen
from their Poisson brackets  on the constraint shell:
\bea
\Delta_{a\bar a}(1,2)\equiv
\left. \left\{ \J_a^{\cH}(Z_1),\J_{\bar a}^{\cH} (Z_2) \right\} 
 \right|_{\J_{{\widehat {gl}}(2)\oplus {\widehat {sl}}(2|1)}=0}& = & 
  K\delta_{a{\bar a}}D\Db \delta (1,2) ,\nn \\
{\overline\Delta}_{{\bar a}a}(1,2)
\equiv\left. \left\{ \J_{\bar a}^{\cH}(Z_1),\J_{a}^{\cH} (Z_2) \right\} 
  \right|_{\J_{{\widehat {gl}}(2)\oplus {\widehat {sl}}(2|1)}=0}& = & 
  -K\delta_{{\bar a}a}\Db D \delta (1,2), \label{dirac1}
\eea
which contain the non--invertible chiral/antichiral projectors $D\Db$ and
$\Db D$ acting on the delta-function (in the above equation, we use the compact 
notation
$\J^{\cH}_a=\left\{h_1,h_2,-j_{12},-j_{34},-j_{35},j_{45}\right\}$ and 
$\J^{\cH}_{\bar a}=\left\{h_{\bar 1}, h_{\bar 1}+h_{\bar 2},j_{21},
j_{43},j_{53},j_{54}\right\}$). Nevertheless, one can
define the following brackets for any supercurrent $\J$ \p{jmn} 
from the N=2 ${\widehat {sl}}(3|2)$ superalgebra:
\bea \label{db}
\left\{ {\cal J}(Z_1) ,{\cal J}(Z_2)  \right\}{}^{*} & \equiv &\left[
\left\{ {\cal J}(1) ,{\cal J}(2)  \right\} -\int\!dZ_3dZ_4\left(
\left\{ {\cal J}(1), \J^{\cH}_a(3)  \right\} 
  {\overline\Delta}^{a{\bar a}}(3,4)
       \left\{\J^{\cH}_{\bar a}(4) ,{\cal J}(2)  \right\}  \right.
           \right. \nn \\
  & & -\left.\left.\left.\left\{ {\cal J}(1), \J^{\cH}_{\bar a}(3) 
\right\} 
        \Delta^{{\bar a}a}(3,4)
       \left\{\J^{\cH}_a(4) ,{\cal J}(2)  \right\}\right)
   \right]\right|_{\J_{{\widehat {gl}}(2)\oplus {\widehat {sl}}(2|1)}=0} \;,
\eea
where we introduced the operators $\Delta^{{\bar a}a}(1,2)$ and
${\overline\Delta}^{{a\bar a}}(1,2)$
which  are required to satisfy the following equations
\bea
&&\int\! dZ_3\Delta_{a{\bar a}}(1,3)\Delta^{{\bar a}b}(3,2) =
  -\delta_a^b D\Db \partial^{-1}\delta (1,2),\;
\int\! dZ_3\Delta^{{\bar b}a}(1,3)\Delta_{a{\bar a}}(3,2)=
  -\delta_{\bar a}^{\bar b} D\Db \partial^{-1}\delta (1,2),\nn \\
&&\int\! dZ_3{\overline\Delta}_{{\bar a}a}(1,3)
       {\overline\Delta}^{a{\bar b}}(3,2) =
  -\delta_{\bar a}^{\bar b} \Db D \partial^{-1}\delta (1,2),\;
\int\! dZ_3{\overline\Delta}^{b{\bar a}}(1,3)
       {\overline\Delta}_{{\bar a}a}(3,2)=
  -\delta_a^b \Db D \partial^{-1}\delta (1,2).\nn\\
&&  \label{xxx}
\eea
The brackets \p{db} possess the main property of the Dirac brackets:
\be
\left\{ \J, \J_{\cH} \right\}^{*} =0, \label{zzz} 
\ee
so we will call them simply Dirac brackets (to check \p{zzz} one should use 
eqs.\p{xxx}, the algebra \p{DD} for the fermionic derivatives and take 
into account that on the constraint shell 
$\J_{{\widehat {gl}}(2)\oplus {\widehat {sl}}(2|1)} =0$  
the Poisson brackets $\left\{ {\cal J}(Z_1), \J^{\cH}_a(Z_2)  \right\}$ become
chiral on the second point while the brackets
$\left\{\J^{\cH}_{\bar a}(Z_1) ,{\cal J}(Z_2)  \right\}$ are antichiral on
the first point). For the case under consideration the explicit solutions
of eqs.\p{xxx} are
\be
 \Delta^{{\bar a}a}(1,2)=-{\overline\Delta}^{a{\bar a}}(1,2)=
  -\frac{1}{K}\delta^{a\bar a} \partial^{-1}\delta(1,2),
\ee
which are defined up to obvious irrelevant terms which do not change 
the brackets \p{db}.
Now we can calculate the Dirac brackets between the matrix superfields 
$F,{\overline F}$ \p{mat2}. They form the following closed non--local algebra
\bea
\left\{ F_{Aa} (Z_1),F_{Bb} (Z_2) \right\}{}^{*}& = & 
    \frac{1}{K}\left( F_{Ba} D\Db \partial^{-1}F_{Ab} -
    (-1)^{d_{a}d_{b}} F_{Ab} D\Db \partial^{-1}F_{Ba}
        \right) \delta (1,2) \; , \nn \\
\left\{ {\overline F}_{aA} (Z_1),{\overline F}_{bB} (Z_2) \right\}{}^{*}& = & 
    \frac{1}{K}\left(  
   (-1)^{d_a d_b}{\overline F}_{bA} \Db D \partial^{-1}{\overline F}_{aB} -
   {\overline F}_{aB} \Db D \partial^{-1}{\overline F}_{bA}
        \right) \delta (1,2) \; ,  \nn \\
\left\{ F_{Aa} (Z_1),{\overline F}_{bB} (Z_2) \right\}{}^{*}& = &
 \left( -K (-1)^{d_b}\delta_{ab}\delta_{AB} D\Db +
    \frac{(-1)^{d_b}}{K}\delta_{ab} F_{Ac}D\Db\partial^{-1}
        {\overline F}_{cB} \right. \nn \\
    & & -\left. \frac{1}{K}\delta_{AB} F_{Ca}D\Db\partial^{-1}
        {\overline F}_{bC}\right) \delta (1,2) .\label{newscanl}
\eea
The algebra \p{newscanl}  
is a particular case of the non--local algebras
proposed in \cite{BKS2}. Let us remark that validity of the
Jacobi identities for it is guaranteed
since we constructed it by applying Dirac procedure
to the local N=2  superalgebra ${\widehat {sl}}(3|2)$.

Now one can easily check that the Hamiltonians \p{h12}
with the covariant derivatives replaced by the fermionic ones \p{DD}
produces the correct flow equations \p{mgnls} if we use the brackets
\p{newscanl}.
Let us stress that generically the non--local superalgebra 
\p{newscanl} is valid for the all cosets of N=2 ${\widehat {sl}}(3|2)$ we
are considering in this section.

\subsection{N=2 
${\widehat {sl}}(3|2)/{\widehat {gl}}(3)\oplus {\widehat {sl}}(2)$ 
coset construction}
The third admissible subalgebra from the list \p{sub1},\p{sub2} 
$\cH_3={\widehat {gl}}(3)\oplus {\widehat {sl}}(2)$ gives rise to the following
decomposition for the supercurrents
\be
\J_{\cH_3}=
\left( \begin{array}{ccc|cc}
 h_{\bar 1} &j_{12} &j_{13} & &  \\ 
 j_{21} & h_{\bar 2}+h_1 &j_{23} & &   \\
 j_{31}&j_{32} &  h_2 &  & \\ \hline
 & &   & h_{\bar 1}+h_1 & j_{45} \\
 &  & & j_{54}& h_{\bar 2}+h_2 
\end{array} \right) , \;
\J_{{\widehat {sl}}(3|2)/{\cH_3}}=
\left( \begin{array}{ccc|cc}
  & & &j_{14} & j_{15} \\ 
  & & &j_{24} & j_{25}    \\ 
  & & &j_{34} & j_{35}      \\ \hline
j_{41} & j_{42} & j_{43} & &     \\
j_{51} & j_{52} & j_{53} & &     
\end{array} \right).
\label{coset3}
\ee
We again arrange the coset supercurrents into
two matrices -- $F$ and ${\overline F}$:
\be
F=\left\{ \begin{array}{cc}
 j_{14} & j_{15} \\
  j_{24} & j_{25} \\
  j_{34} & j_{35} 
\end{array}
 \right\} \quad ,
{\overline F}=\left\{
\begin{array}{ccc}
j_{41}& j_{42} & j_{52}\\
j_{51}& j_{52} & j_{53} 
\end{array} \right\} . \label{mat3}
\ee
The covariant derivatives for the coset currents \p{mat3} are
\bea
\cDb j_{14} & = & \Db j_{14}+\frac{1}{K}\left( j_{14}h_{\bar 1}+
 j_{15}j_{54}\right) \; , \;
\cDb j_{15}  =  \Db j_{15}+\frac{1}{K}\left( j_{15}h_{\bar 1}+
  j_{15}h_{\bar 2} \right)\; , \nn \\
\cDb j_{24} & = & \Db j_{24}+\frac{1}{K}\left( j_{14}j_{21}+
  j_{25}j_{54}\right) \; , \;
\cDb j_{34}  =  \Db j_{34}-\frac{1}{K}\left( j_{34}h_{\bar 2}-
 j_{14}j_{31}-j_{24}j_{32}-j_{35}j_{54} \right)\; , \nn \\
\cDb j_{25} & = & \Db j_{25}+\frac{1}{K}\left( j_{25}h_{\bar 2}+
  j_{15}j_{21} \right)\; , \;
\cDb j_{35}  =  \Db j_{35}+\frac{1}{K}\left( j_{15}j_{31}+
  j_{25}j_{32}\right) \; , \nn \\
\cD j_{41} & = & D j_{41}+\frac{1}{K} j_{45}j_{51} \; , \;
\cD j_{42}  =  D j_{42}+\frac{1}{K}\left( h_1 j_{42}+
 j_{12}j_{41}+j_{45}j_{52}\right) \; , \nn \\
\cD j_{43} & = & D j_{43}+\frac{1}{K} \left( h_1 j_{43}+ h_2 j_{43}+
  j_{13}j_{41}+j_{23}j_{42}+j_{45}j_{53} \right) \; , \;
\cD j_{51}  =  D j_{51}-\frac{1}{K} h_1 j_{51}\;, \nn \\
\cD j_{52} & = & D j_{52}+\frac{1}{K} j_{12} j_{51}\;, \;
\cD j_{53}  =  D j_{53}+\frac{1}{K} \left( h_2 j_{53}+j_{13}j_{51}+
   j_{23} j_{52}\right) \; . \label{cd3}
\eea
Again one can check that the Hamiltonians defined in \p{h12}, applied to 
this case,
produce the first and the second flows \p{mgnls}
of the N=2 $(3|2,0)$--MGNLS hierarchy.

The coset construction we considered in this subsection contains only bosonic 
superfields which we arranged in a $(3\times 2)$ matrix $F$ and a 
$(2\times 3)$ matrix $\overline F$ \p{mat3}. The Dirac brackets for them
are given
by the non--local superalgebra \p{newscanl}. But here
comes an important observation. There exists another non--local superalgebra
of the type \p{newscanl} with the same number of degrees of
freedom -- 6 chiral and 6 antichiral bosonic superfields --
but with a different matrix arrangement -- a $(2\times 3)$ matrix $F_1$ and a
$(3\times 2)$ matrix $\overline F_1$. The latter system describes
a different hierarchy -- the N=2 $(2|3,0)$--MGNLS one. 
One can verify that actually these two algebras
coincide after the redefinition $F_1=iF^T,{\overline F}_1=i{\overline F}{}^T$
(the superscript $T$ denotes matrix transposition).
Therefore, in the framework of the same N=2 coset construction 
${\widehat {sl}}(3|2)/{\widehat {gl}}(3)\oplus {\widehat {sl}}(2)$
with the matrix superfields $F, \overline F$ \p{mat3} there is room for two
different hierarchies,
 the N=2 $(3|2,0)$--MGNLS and
the N=2 $(2|3,0)$--MGNLS hierarchies. The Hamiltonians and the flows
corresponding to the last hierarchy can be obtained from eqs.\p{h12} and
\p{mgnls}, respectively, via the following substitutions:
\be
F\rightarrow i F^T,\quad\quad \overline F \rightarrow 
i \overline F^T, \label{trans1}
\ee
where $i$ is the imaginary unit.

What we just illustrated in the above example is actually completely
general. One can show that the Hamiltonian structure for the 
N=2 $(k|l,0)$--MGNLS hierarchy is the same as for the N=2 $(l|k,0)$--MGNLS
one. One Hamiltonian structure is mapped to the other by the 
transformation \p{trans1}.
Similarly the 
N=2 $(k|0,m)$--MGNLS and the N=2 $(m|0,k)$--MGNLS hierarchies, which
involve pure fermionic matrix superfields, possess
the same second Hamiltonian structure. In this case one Hamiltonian structure
is mapped to the other by the transformation $F \rightarrow F^T$ and
$\overline F \rightarrow \overline F^T$.   

So one can state that to any
non--local superalgebra \p{newscanl} with either only bosonic or only fermionic
matrices, there correspond two different N=2 MGNLS hierarchies.
It is natural to remark the analogy with the case of the N=2 $W_S$ 
superalgebra which is the second Hamiltonian structure for three different
supersymmetric KdV--type hierarchies.

\subsection{N=2 ${\widehat {sl}}(3|2)/{\widehat {gl}}(3|1)$ coset 
construction}
The last case from the list \p{sub1},\p{sub2} for the admissible N=2 
${\widehat {sl}}(3|2)$ subalgebras
is $\cH_4= {\widehat {gl}}(3|1)$:
\be
\J_{\cH_4}=
\left( \begin{array}{cccc|c}
 h_{\bar 1} &j_{12} &j_{13} &j_{14} &  \\ 
 j_{21} & h_{\bar 2}+h_1 &j_{23} &j_{24} &   \\
 j_{31}&j_{32} &  h_2 & j_{34} & \\ 
 j_{41}&j_{42} &j_{43}  & h_{\bar 1}+h_1 &  \\ \hline
 &  & & & h_{\bar 2}+h_2 
\end{array} \right), \;
\J_{{\widehat {sl}}(3|2)/{\cH_4}}=
\left( \begin{array}{cccc|c}
  & & & & j_{15} \\ 
  & & & & j_{25}    \\ 
  & & & & j_{35}      \\ 
  & & & & j_{45}    \\ \hline
j_{51} & j_{52} & j_{53} & j_{54} &     
\end{array} \right).
\label{coset4}
\ee
Among the  coset supercurrents only $j_{45}$ and $j_{54}$ are fermionic while 
the remaining ones are bosonic.
We arrange these supercurrents into
$F$ and ${\overline F}$:
\be
F=i\left\{ j_{15}, j_{25},  j_{35} , j_{45} \right\}, \quad 
{\overline F}=i\left\{
\begin{array}{c}
j_{51}\\ j_{52} \\ j_{53}\\ j_{54}
\end{array} \right\} . \label{mat4}
\ee
The covariant derivatives for the coset currents \p{mat4} 
read:
\bea
\cDb j_{15} & = & \Db j_{15}+\frac{1}{K}\left( j_{15}h_{\bar 1}+
  j_{15}h_{\bar 2} \right)\; , \;
\cDb j_{25}  =  \Db j_{25}+\frac{1}{K}\left( j_{25}h_{\bar 2}+
  j_{15}j_{21} \right)\; , \nn \\
\cDb j_{35} & = & \Db j_{35}+\frac{1}{K}\left( j_{15}j_{31}+
  j_{25}j_{32}\right) \; , \;
\cDb j_{45}  =  \Db j_{45}-\frac{1}{K}\left( j_{45}h_{\bar 2}-
  j_{15}j_{41}-j_{25}j_{42}-j_{35}j_{43} \right)\; , \nn \\
\cD j_{51} & = & D j_{51}-\frac{1}{K} h_1 j_{51}\;, \;
\cD j_{53}  =  D j_{53}+\frac{1}{K} \left( h_2 j_{53}+j_{13}j_{51}+
   j_{23} j_{52}\right) \; , \nn \\
\cD j_{52} & = & D j_{52}+\frac{1}{K} j_{12} j_{51}\;, \;
\cD j_{54}  =  D j_{54}-\frac{1}{K} \left( h_1 j_{54}+j_{14}j_{51}+
   j_{24} j_{52}+j_{34}j_{53}\right) \; . \label{cd4}
\eea
For this coset construction the Hamiltonians \p{h12} again produce 
via \p{mgnls} the flows
which correspond to the N=2 $(1|3,1)$--MGNLS hierarchy.

\sect{Conclusion and outlook}

In the previous section we explicitly demonstrated that all the subalgebras 
of the N=2 ${\widehat {sl}}(3|2)$ superalgebra from the list 
\p{sub1},\p{sub2}, i.e. the admissible ones, give rise via coset construction
to N=2 MGNLS hierarchies (GNLS ones, when the matrix superfields $F$ and 
$\overline F$ are simple row and column matrices, respectively). 
We have checked that 
the same is true for the two admissible subalgebras 
of N=2 ${\widehat {sl}}(2|1)$. 
These examples suggest that for each N=2 
${\widehat {sl}}(n|n-1)$ superalgebra its admissible subalgebras 
\p{sub1},\p{sub2} give rise to the N=2 MGNLS 
hierarchies and the corresponding Dirac reductions of the local
second Hamiltonian structure (N=2 ${\widehat {sl}}(n|n-1)$)
reproduce the corresponding non--local second Hamiltonian 
structures of the N=2 MGNLS hierarchies, \cite{BKS2}.

We would like to point out that the coset construction can be applied to any
N=2 subalgebra which appears as addenda in the list \p{sub1},\p{sub2}, e.g. to
${\widehat {gl}}(2n|2m)$ and ${\widehat {sl}}(2n+1|2m)$.
For example, for the case of N=2  ${\widehat {sl}}(2n+1)$ superalgebras 
this procedure leads to the 
N=2 GMNLS hierarchies with pure fermionic matrix superfields. 
We have explicitly checked this 
for the cases of N=2  ${\widehat {sl}}(3)$ and ${\widehat {sl}}(5)$ 
superalgebras. The coset constructions
related to the N=2 ${\widehat {sl}}(2n+1)$ superalgebra involve only
fermionic matrix superfields and we can construct two different integrable 
hierarchies for each admissible subalgebra -- the N=2 $(s|0,2n+1-s)$ and the 
N=2 $(2n+1-s|0,s)$--MGNLS ones. This is in fact a particular case of what was 
remarked at the end of subsection 3.3.

Now we should recall the original motivation of the present work, namely
understanding the origin of the non--locality of the second Hamiltonian 
structure of all N=2 MGNLS hierarchies.
In this sense the main result of our paper is that we have been able to 
`localize' these non--local Hamiltonian structures. In fact, as we have 
already remarked, the
original superalgebras we started from are local and we could make a list
of admissible subalgebras, thus
providing a sort of classification for the N=2 MGNLS (GNLS) hierarchies.
Moreover these superalgebras, which
localize the second Hamiltonian structures for N=2 MGNLS hierarchies, can
give rise to new hierarchies with extended sets of superfields.
For the case N=2 $gl(2)$ this was demonstrated in \cite{IKT}.

Let us conclude our letter with a few remarks. 
One of the interesting consequences of our approach is the fact that some
of the N=2 MGNLS (GNLS) hierarchies are more "fundamental" than others. 
Indeed, in the
framework of coset construction we were able to reproduce the matrix hierarchies
only  for some specific relations between the numbers of rows and columns --
precisely, the
supermatrix $F$ is generically defined as an 
$s\times (2n-1-s)$ rectangular matrix,
with $s=1,\ldots ,2n-2$, for the case of N=2 $\widehat {sl}(n|n-1)$. 
Of course, the resulting systems admit further reductions to the 
$s_1\times n_1$ supermatrix, with $s_1\leq s, n_1\leq 2n-1-s$ , giving
rise to all the N=2 MGNLS hierarchies \cite{BKS2}, but it might not be possible 
to
construct these "secondary" reduced cases directly via the coset approach.
It would be interesting to understand the reason of this `fundamental' role.
Another mysterious fact is that the number $4sn-2s(s+1)$ of the N=2
superfields in all the admissible coset constructions is divisible by
4, thus 
providing the necessary condition for the existence of a hidden $N=4$
supersymmetry. In \cite{BS} it is has been verified that actually some of them 
do possess a hidden $N=4$ supersymmetry.

Finally we would like to mention the possibility to use the coset 
approach for the superalgebras which come as a result of Hamiltonian
reduction applied to the N=2 ${\widehat {sl}}(n|n-1)$ superalgebra \cite{AIS} 
(e.g. N=2 $W_S$ superalgebras, etc.). The first example of such 
construction has been elaborated in \cite{BK}.

\section*{Acknowledgements}
We would like to thank P.P. Kulish for his interest in this paper.
S.K. is grateful to E.A. Ivanov and F. Toppan for many useful discussions.
S.K. and A.S. thank SISSA  for the hospitality 
during the final stage of this research. This investigation has been
supported in part by grants RFBR 96-02-17634, RFBR-DFG
96-02-00180, INTAS-93-127 ext., INTAS-93-1038 and by the EC TMR
Programme Grant FMRX-CT96-0012.

\end{document}